\documentclass[sigconf]{acmart}
\usepackage{graphicx}
\usepackage{url}
\usepackage{siunitx}
\usepackage{pifont}
\usepackage{booktabs}
\usepackage{microtype}
\usepackage{multicol}
\definecolor{darkgreen}{RGB}{0,100,0}
\newcommand{\cmark}{\color{darkgreen}\ding{51}}
\newcommand{\xmark}{\color{red}\ding{55}}

\AtBeginDocument{%
  \providecommand\BibTeX{{%
    \normalfont B\kern-0.5em{\scshape i\kern-0.25em b}\kern-0.8em\TeX}}}

\copyrightyear{2025}
\acmYear{2025}
\setcopyright{rightsretained}
\acmConference[UbiComp Companion '25]{Companion of the 2025 ACM International Joint Conference on Pervasive and Ubiquitous Computing}{October 14--16, 2025}{Espoo, Finland} \acmBooktitle{Companion of the 2025 ACM International Joint Conference on Pervasive and Ubiquitous Computing (UbiComp Companion '25), October 14--16, 2025, Espoo, Finland} 
\begin{document}

\title{Push or Light: Nudging Standing to Break Prolonged Sitting}

\author{Sohshi Yoshida}
\orcid{}
\affiliation{%
  \institution{Osaka Metropolitan University}
  \city{Osaka}
  \country{Japan}
}
\email{sm22112v@st.omu.ac.jp}
\authornote{These authors contributed equally to this research.}

\author{Ko Watanabe}
\orcid{0000-0003-0252-1785}
\affiliation{%
  \institution{DFKI}
  \city{Kaiserslautern}
  \country{Germany}
}
\email{ko.watanabe@dfki.de}
\authornotemark[1]

\author{Andreas Dengel}
\orcid{0000-0002-6100-8255}
\affiliation{%
  \institution{RPTU Kaiserslautern-Landau {\&} DFKI}
  \city{Kaiserslautern}
  \country{Germany}
}
\email{andreas.dengel@dfki.de}

\author{Shoya Ishimaru}
\orcid{0000-0002-5374-1510}
\affiliation{%
  \institution{Osaka Metropolitan University}
  \institution{{\&} DFKI Lab Japan}
  \city{Osaka}
  \country{Japan}
} 
\email{ishimaru@omu.ac.jp}

\author{Shingo Ata}
\orcid{0009-0006-4013-3563}
\affiliation{%
  \institution{Osaka Metropolitan University}
  \city{Osaka}
  \country{Japan}
}
\email{ata@omu.ac.jp}

\author{Manato Fujimoto}
\orcid{0000-0002-6171-5697}
\affiliation{%
  \institution{Osaka Metropolitan University}
  \city{Osaka}
  \country{Japan}\\
  \institution{RIKEN Center for Advanced Intelligence Project AIP}
  \city{Tokyo}
  \country{Japan}
}
\email{manato@omu.ac.jp}

\renewcommand{\shortauthors}{Yoshida and Watanabe et al.} 

\begin{abstract}
Prolonged sitting is a health risk leading to metabolic and cardiovascular diseases. 
To combat this, various ``nudging'' strategies encourage stand-ups. 
Behavior change triggers use explicit prompts such as smartphone push notifications or light controls. 
However, comparisons of the effects of such interactions, discomfort, and user context have not yet been performed.
The present study evaluated these methods in a mixed design experiment with 15 college students.
Three intervention methods (none, push notifications, and light dimming) and three user task contexts (computer work, video calls, and reading) were tested.
The frequency of standing up and comfort were assessed after each ten-minute session.
Results showed that dimming resulted in slightly more breaks (1.4 ± 1.55) than push notification (1.2 ± 1.08), but caused discomfort for 66.7\% of participants, compared to 20\% for notification.
The results were influenced by task context. Dimming was most effective during video calls and reading, while push notifications were more effective during computer work.
These findings suggest adaptive nudging systems should tailor interventions based on context and individual preferences.
\end{abstract}


\begin{teaserfigure}
  \centering 
  \includegraphics[width=0.75\textwidth]{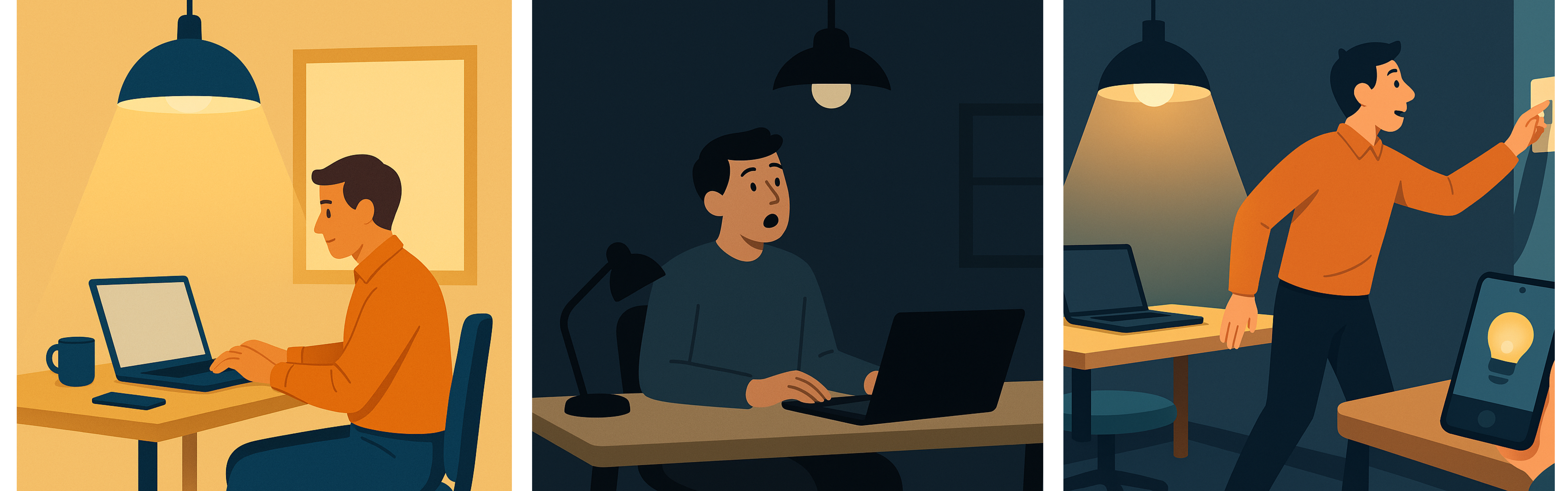}
  \caption{Overview of our study: Evaluate the impact of the ambient light control and the behavioral change.}
  \Description{This image highlights the overview of the target intervention and impact of the behavior change.}
  \label{fig:teaser}
\end{teaserfigure}


\maketitle

\begin{table*}[t]
  \centering
  \renewcommand{\arraystretch}{1.0}
  \caption{Position of our work against existing behavior change intervention (Shikakeology). \cmark = supported, \xmark = not supported.}
  \vspace{0.5em}
  \resizebox{\textwidth}{!}{
  \begin{tabular}{lccccll}
    \toprule
    \textbf{Reference}                         & \textbf{Modality} & \textbf{Stand Up} & \textbf{Light} & \textbf{Context}   & \textbf{Cue / Mechanism}                  & \textbf{Target Detail}   \\
    \midrule
    \citet{lyons2024defining}                  & Wearable          & \cmark            & \xmark         & \xmark             & Apple Watch Stand-Hour notification       & Encourage stand breaks   \\
    \citet{ailneni2019influence}               & Wearable          & \cmark            & \xmark         & \xmark             & Apple Watch push notification             & Encourage stand up       \\
    \citet{fun2014fun}                         & Ambient           & \xmark            & \xmark         & \xmark             & Piano-sound staircase                     & Encourage stair use      \\
    \citet{jafarinaimi2005breakaway}           & Ambient           & \cmark            & \xmark         & \xmark             & Slouching desk sculpture                  & Take a break             \\
    \citet{fortmann2013movelamp}               & Ambient           & \cmark            & \cmark         & \xmark             & Battery-metaphor color-gradient lamp      & Increase steps \& breaks \\
    \citet{mateevitsi2014healthbar}            & Ambient           & \cmark            & \cmark         & \xmark             & Light metaphor of prolonged sitting       & Health coach \& Encourage breaks\\
    \citet{Yamanishi2023SunsetLighting}        & Ambient           & \cmark            & \cmark         & \xmark             & Gradual color shift in restroom lights    & Shorten restroom stay    \\
    \midrule
    \textbf{This work}                         & Ambient           & \cmark            & \cmark         & \cmark             & Push Notification vs. Dimming of lights   & Encourage stand up       \\
    \bottomrule
  \end{tabular}
  }
  \label{tab:research-position}
\end{table*}

\section{Introduction}
\label{sec:introduction}
Prolonged sedentary behavior has been shown to negatively impact peripheral blood circulation and metabolic function negatively, increasing the risk of cardiovascular disease and all-cause mortality~\cite{koyama2021effect}.
To address these health risks, recent studies have focused on using wearable devices~\cite{okoshi2015reducing, tanaka2024concentration} to reduce extended sitting periods. For instance, the smart watch provides a ``Stand Notification'' to prompt users to stand after sitting for too long~\cite{lyons2024defining}. While a field study found that 74\% of users responded by standing~\cite{ailneni2019influence}, about 26\% ignored the notification, highlighting limitations such as the need for continuous device use and potential disruption to users' routines.

In contrast, the field of \emph{Shikakeology}~\cite{matsumura2013shikake} explores subtle environmental interventions that naturally encourage behavioral change without explicit user engagement. Examples include a piano-key staircase that promotes stair use through sound, playful rubbish bins that enhance waste disposal, a urinal fly decal that minimizes splashing~\cite{fun2014fun}, and restroom lighting that changes color to reduce occupancy time~\cite{Yamanishi2023SunsetLighting}. These interventions blend into daily life, requiring no active user participation.

Building on this concept, our study investigates whether ambient light control can act as a non-intrusive signal to encourage office workers to stand. \autoref{fig:teaser} shows the overview of our study. We utilize Philips Hue lights~\cite{PhilipsHueBridge} and illuminance sensors to dim the workspace lighting after prolonged sitting. Our research aims to answer the following questions:

\begin{itemize}
  \item[\textbf{RQ1}] Can dynamic light control effectively prompt individuals to stand after extended periods of sitting?
  \item[\textbf{RQ2}] What user conditions enhance the effectiveness of light control in prompting standing behavior?
\end{itemize}

\section{Related Work}
Table~\ref{tab:research-position} offers an overview of the current research landscape, positioning our study within it. 
This section explores various strategies developed to tackle excessive sedentary behavior, focusing on interventions using wearable technology and Shikakeology principles.

\subsection{Wearable Technology-Based Interventions}
Interventions using wearable technology have gained significant attention for addressing sedentary lifestyles. 
Devices like the smart watch track prolonged inactivity and prompt users to stand through push notifications~\cite{lyons2024defining}. 
A study assessing the ``Stand Notification'' feature found that 74\% of users responded by standing up~\cite{ailneni2019influence}. 
However, 26\% of users did not respond, highlighting the variability in effectiveness among different users.

A significant limitation of wearable-based interventions is their reliance on continuous device usage and user attention to notifications. 
Depending on explicit prompts, these interventions may disrupt natural behavior patterns, potentially leading to resistance or notification fatigue. 
This highlights the need for more seamless, non-intrusive methods to promote healthy behaviors.

\subsection{Ambient Technology-Based Interventions}
Unlike wearable technologies, the field of ``Shikakeology''~\cite{matsumura2013shikake} provides valuable insights into ambient technology-based interventions.
\emph{Shikakeology} focuses on design and analyze interventions that naturally induce behavioral change without external pressure or explicit instructions.
An example of successful Shikakeology-based interventions includes a project that installed pressure sensors on stair steps, which played piano sounds when stepped on, resulting in a 66\% increase in stair usage compared to the baseline~\cite{fun2014fun}.

\citet{jafarinaimi2005breakaway} introduced \emph{BreakAway}, a study designed to encourage office workers to stand through interaction. The concept involves using a mesh-based seat sensor to detect the participant's sitting status. Based on this sensor data, the study visualizes the seating status through a sculpture. After 60 and 90 minutes of sitting, the sculpture adopts progressively slouched poses. When the chair is unoccupied, the sculpture returns upright over ten minutes.

\begin{figure*}[t!]
  \centering
  \includegraphics[width=\linewidth]{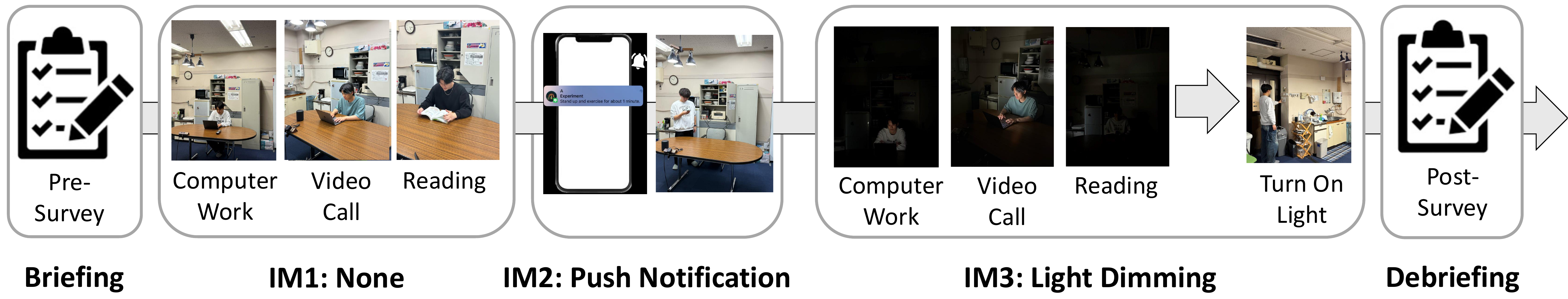}
  \caption{Overview of the experiment procedures. Participants assigned different task context (TC) and experience three different intervention modes (IM).}
  \label{fig:system}
\end{figure*}

\begin{figure}[t!]
  \centering
  \includegraphics[width=\linewidth]{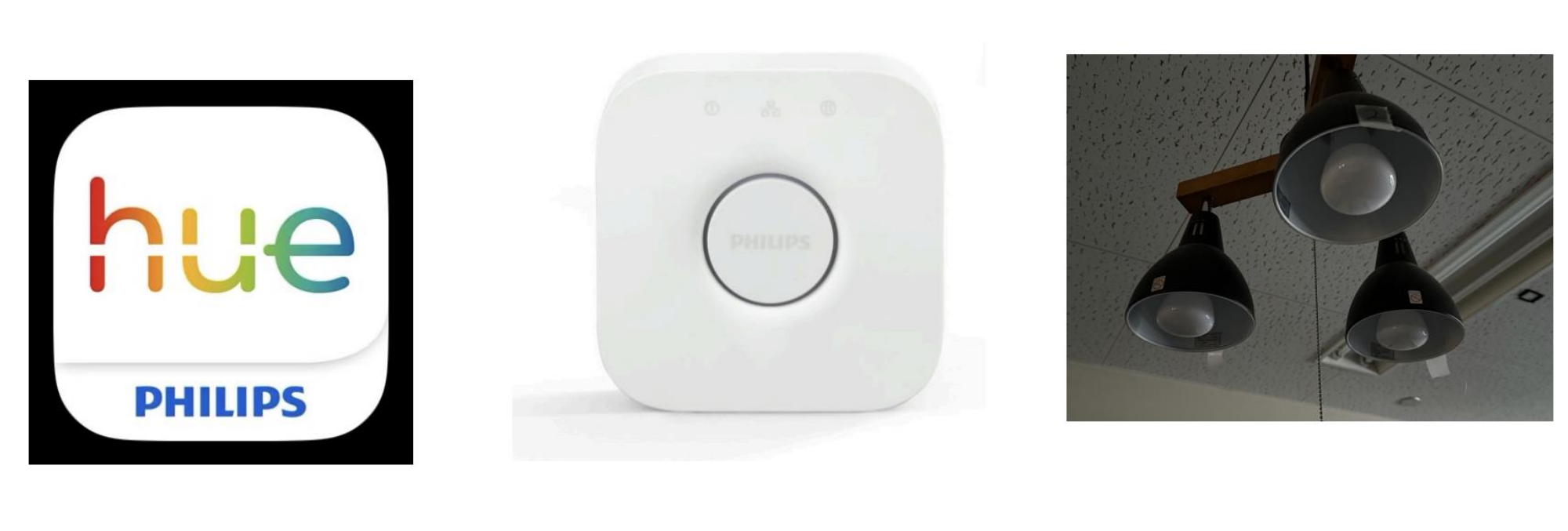}
  \caption{Equipment for lighting control. The Philips Hue application is connected to the light bulb, which can be turned on and off via the internet.}
  \label{fig:sensor_information}
\end{figure}

\citet{fortmann2013movelamp} proposed \emph{MoveLamp}, which uses a mobile phone pedometer as an input to display a color gradient lamp located on the office desk. 
The display changes color based on the number of steps taken. The study successfully increased the steps taken by 57.48\% and movement frequency by 21.64\% compared to the control condition.

\citet{mateevitsi2014healthbar} introduce \emph{HealthBar}, which supports cognitive augmentation of participants by representing a health gauge using an ambient persuasive light device.
\emph{HealthBar} indicates complete discharge after 45 minutes, so participants encourage to stand and rest before being discharged.

\citet{Yamanishi2023SunsetLighting} introduces intervention using lighting-based mechanisms in public restrooms to address prolonged use. 
The intervention reduced restroom usage by approximately 60\% by gradually changing the lighting color based on occupancy duration. 
This study demonstrate how subtle, context-sensitive environmental modifications can guide human behavior.

\subsection{Position of Our Work}
Our study sets itself apart from previous research by concentrating on ambient technology interventions, specifically through light control, to combat sedentary behavior. As illustrated in \autoref{tab:research-position}, our approach is closely aligned with the work of \citet{mateevitsi2014healthbar}, \citet{fortmann2013movelamp} or \citet{Yamanishi2023SunsetLighting}. However, our study contributes to comparing push notifications and focuses on the context of the user's activity during interventions, aiming to understand how different activities influence the effectiveness of interventions.

\section{Methodology}
This section outlines the methodology employed in our study, detailing the experimental design, system design, and questionnaires.

\subsection{Experimental Design}
Figure~\ref{fig:system} shows the experimental setup. We conducted a controlled laboratory study to compare three intervention modes—\textit{None}, \textit{Push Notification}, and \textit{Ambient Light Dimming}—concerning their ability to break prolonged sitting. The experiment followed a both within-subjects and between-subjects design: \textbf{intervention mode} was a within-subjects factor (all participants encountered the same three ten-minute phases in fixed order), whereas \textbf{task context} (\emph{computer work}, \emph{video call}, and \emph{reading}) was a between-subjects factor (five participants per context).

\subsubsection{Intervention Mode}
\label{sec:system}
We set up a windowless room with three Philips Hue \textit{White \& Color Ambiance} bulbs, all connected to a Hue Bridge. 
Figure~\ref{fig:sensor_information} illustrates the lighting control equipment. Below are the three intervention modes (IM1--IM3) along with their respective control conditions:

\begin{enumerate}
  \item[IM1] \textbf{None:} Participants engaged in tasks using a laptop computer without any intervention.
  \item[IM2] \textbf{Push Notification:} Participants received smartphone messages prompting them to ``\emph{Stand up and stretch for one minute}.''
  \item[IM3] \textbf{Light Dimming:} A Raspberry Pi 4 ran a Python script to send REST-API calls for dimming the lights.
\end{enumerate}

Throughout implementing these intervention modes, we manually tracked the number of times participants stood up.

\subsubsection{Task Context}
Participants were randomly assigned to one of three realistic desk-based task contexts (TC1--TC3):

\begin{enumerate}
  \item[TC1] \textbf{Computer Work:} Work freely with a laptop computer.
  \item[TC2] \textbf{Video Call:} Conduct a mock interview over google meet.
  \item[TC3] \textbf{Reading:} Read a printed book/material.
\end{enumerate}

By separating the task contexts, we can isolate the effect of the intervention mode on the behavior change.
These three contexts were also chosen to cover different dependencies on ambient lighting and varying levels of attentional demand.
The DFKI Ethics Committee has approved the experimental setting.

\subsection{Questionnaire}
This section describes the pre-/post-study questionnaire used to collect participant data.

\subsubsection{Pre-Study Questionnaire}
\label{sec:pre-questionaire}
In the pre-study questionnaire, we asked participants to provide demographic information such as age, gender, nationality, and occupation.
This information was used to see how individual characteristics influence the response to light control interventions.

\subsubsection{Post-Study Questionnaire}
\label{sec:post-questionaire}
In the post-study questionnaire, we received feedback from participants about the intervention modes.
We asked the following questions:

\begin{enumerate}
  \item[Q1] Did you find the smartphone notifications uncomfortable?
  \item[Q2] Did you find the light dimming uncomfortable?
  \item[Q3] Which intervention made you stand up more naturally: the smartphone notification or the light dimming?
  \item[Q4] Do you regularly receive notifications on your smartphone or smartwatch reminding you to stand up?
  \item[Q5] If you answered "yes" to Q4, do you typically stand up when prompted by these notifications?
  \item[Q6] Please feel free to share any concerns or thoughts.
\end{enumerate}

These subjective questionnaires are used to assess the effect of the intervention modes on the participants' comfort and behavior change.

\section{Data Collection}
In this section, we describe the participants' details and the experimental procedure.
 
\subsection{Participants}
\label{sec:participants}
Fifteen full-time Japanese university students (average age $22.1\pm1.9$ years; 14 male, one female) were recruited via campus mailing lists. 
Participants were Japanese nationals.
All participants provided written informed consent, and the data obtained from the participants were used solely for analytical purposes.

\subsection{Experiment Procedure}
In this experiment, participants follow the timeline below.
The task context varies among participants and was initially explained in the briefing session.
The experiment takes approximately 45 minutes.

\begin{enumerate}
  \item \textbf{Briefing} (10 minutes): Participants first get instructions for the experiment. Once confirmed, participants fill out the consent form and pre-questionnaire.
  \item \textbf{None (IM1)} (10 minutes): Normal room lighting with no intervention.
  \item \textbf{Notification (IM2)} (10 minutes): Every three minutes, a smartphone notification was delivered.
  \item \textbf{Light Dimming (IM3)} (10 minutes): Every three minutes, simultaneous dimming of all Hue bulbs, users could restore light manually using the button on the wall at any time.
  \item \textbf{Debriefing} (5 minutes): Fill out the post-questionnaire.
\end{enumerate}

Participants were told to opt out of the experiment at anytime.

\section{Results and Discussion}
In this section, we answer the research questions mentioned in the Introduction (Section~\ref{sec:introduction}).

\begin{figure}[t!]
    \centering
    \includegraphics[width=0.99\linewidth]{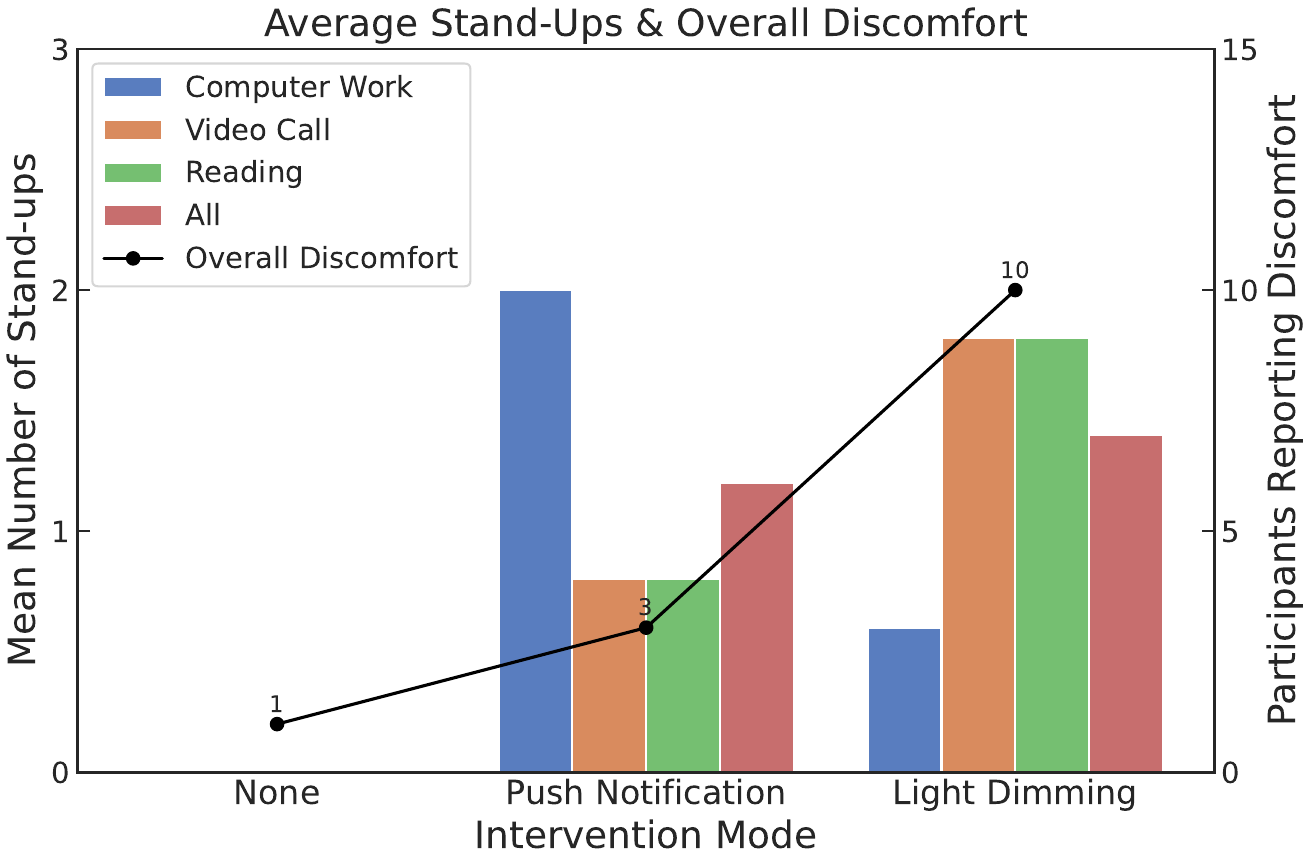}
    \caption{Average number of stand-ups per participant task context and the overall discomfort reported, categorized by intervention mode.}
    \label{fig:perceived_discomfort}
\end{figure}

\subsection{Push Notification \textit{vs.} Light Dimming}

To address \textbf{RQ1}, we compared how often participants stood up under the two intervention modes.  
\autoref{fig:perceived_discomfort} displays, for every task context and the pooled sample, the \emph{mean number of stand-ups} (bars, left axis) together with the \emph{number of participants who found the intervention uncomfortable} (line, right axis).

Participants stood up on average $1.2\pm1.08$ times during the \textit{Push~Notification} phase and $1.4\pm1.55$ times during the \textit{Light~Dimming} phase, indicating that the ambient light cue elicited slightly more activity.  
At the same time, \textit{Light Dimming} generated markedly more discomfort reports (10 of 15) than \textit{Push Notification} (3 of 15).

These results reveal a clear trade-off between \emph{effectiveness} and \emph{comfort}: the stronger, room-wide light cue prompted more stand-ups but was also perceived as more annoying. In other words, a modest level of discomfort catalyzes behaviour change, whereas the subtler push notification was better tolerated but less effective.

\begin{figure}[t!]
  \centering
  \includegraphics[width=0.8\linewidth]{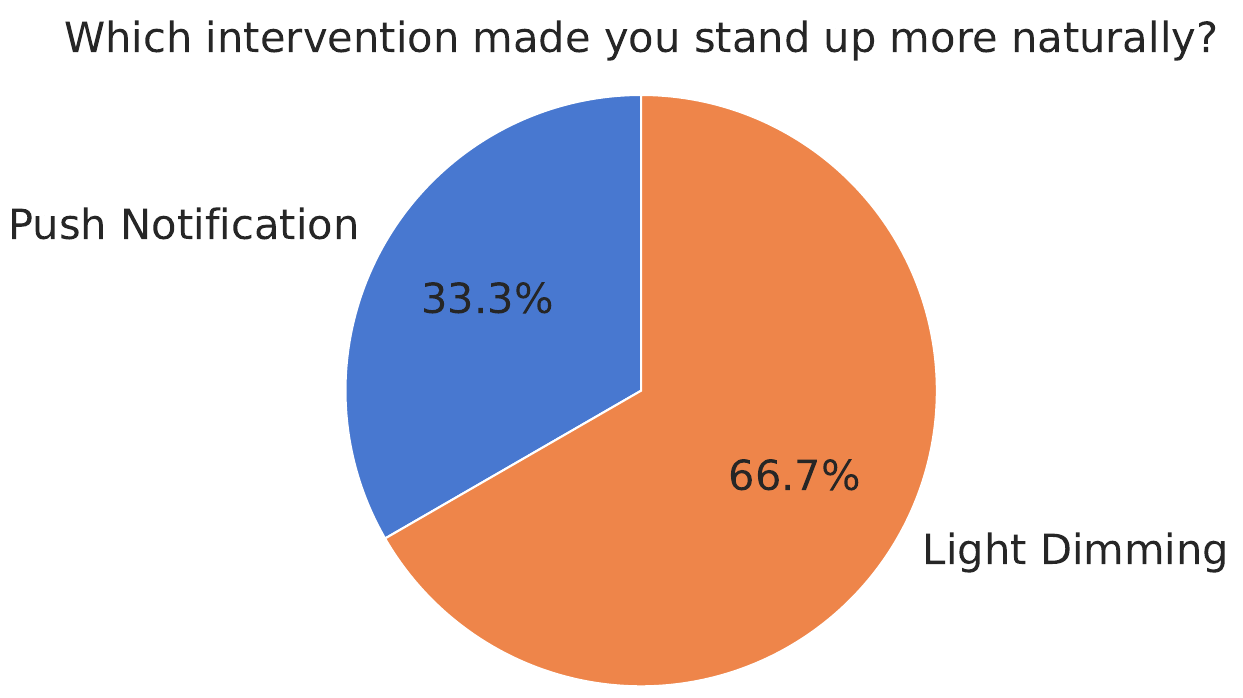}
  \caption{Participants' intervention preferences for stand up, categorized by intervention mode.}
  \label{fig:intervention_preference}
\end{figure}

\autoref{fig:intervention_preference} also shows the participants' intervention preferences for stand up, categorized by intervention mode.
The figure shows that the light dimming intervention was preferred more in natural settings than the push notification intervention.

\subsection{Task Context \textit{vs.} Intervention Mode}
To address \textbf{RQ2}, we compared the average number of stand-ups per participant task context and the overall discomfort reported, categorized by intervention mode.
From \autoref{fig:perceived_discomfort}, we can see that the light dimming intervention enhances standing up for video call and reading tasks.
Meanwhile, the push notification intervention was more effective for computer work.
Depending on the task context, the participants' behavior changed.

The result indicates that activity or context recognition is significant for performing different intervention modes to encourage stand up.
One participant reported in the post-study questionnaire that ``\emph{The computer work did not feel like getting up much because it could work in the dark}''.
This subjective feedback indicates that the light dimming intervention is only effective when the participants cannot work in a dark environment.

\section{Conclusion}
This study explored the effectiveness of ambient light control as a non-intrusive intervention to encourage standing behavior for health. 
Our findings indicate that light dimming can serve as a more effective prompt for standing than traditional push notifications, particularly during tasks that rely on ambient lighting, such as video calls and reading.
However, the increased effectiveness of light dimming came with a trade-off in comfort.
While ambient light control can be a powerful tool for behavior change, its implementation should consider user comfort and task context to optimize its effectiveness. 
Future work could explore adaptive systems that tailor interventions based on real-time activity recognition to balance user comfort and success in behavior changes.

\begin{acks}
This work was supported in part by the Japan Society for the Promotion of Science, Grants-in-Aid for Scientific Research number JP24K02934.
Also, GPT-o3 supported the writing of this manuscript by generating images for the teaser image and sentences.
Human check and remodification are involved after each generation.
\end{acks}

\bibliographystyle{ACM-Reference-Format}
\bibliography{main}
\end{document}